\documentclass[11pt]{article}
\usepackage{graphicx} 
\begin{document}

\def \d {{\rm d}}
\def \U {{\cal U}}
\def \V {{\cal V}}
\def \H {{\cal H}}
\def \M {{\cal M}}
\def \A {{\bf A}}
\def \N {{\bf N}}
\def \T {{\bf T}}

\newcommand{\be}{\begin{equation}}
\newcommand{\ee}{\end{equation}}
\newcommand{\beqn}{\begin{eqnarray}}
\newcommand{\eeqn}{\end{eqnarray}}

\title{Symmetries and geodesics in (anti--)de~Sitter spacetimes with nonexpanding
impulsive waves}

\author{J. Podolsk\'y\thanks{E--mail: {\tt podolsky@mbox.troja.mff.cuni.cz}}
\\
\\ Institute of Theoretical Physics, Charles University,\\
V Hole\v{s}ovi\v{c}k\'ach 2, 18000 Prague 8, Czech Republic.\\
\\
and M. Ortaggio\thanks{E--mail: {\tt ortaggio@science.unitn.it}} \\ \\
Dipartimento di Fisica, Universit\`a degli Studi di Trento,  \\
and INFN, Gruppo Collegato di Trento, 38050 Povo (Trento), Italy. \\  }

\date{\today}

\maketitle
\begin{abstract}
We consider a class of exact solutions which represent nonexpanding
impulsive waves in backgrounds with nonzero cosmological constant.
Using a convenient 5-dimensional formalism it is shown that these
spacetimes admit at least three global Killing vector fields. The same geometrical
approach enables us to find all geodesics in a simple
explicit form and describe the effect of impulsive waves on test particles.
Timelike geodesics in the axially-symmetric Hotta-Tanaka spacetime
are studied in detail. It is also demonstrated that for vanishing
cosmological constant, the symmetries and geodesics reduce to those for
well-known impulsive {\it pp\,}-waves.


\end{abstract}

\section{Introduction}

Impulsive plane waves in Minkowski space  have been studied for decades.
These can naturally be understood as a limiting case of sandwich
{\it pp\,}-waves \cite{BPR59} with wave-profiles approaching the Dirac
delta distribution. Furthermore, Aichelburg and Sexl \cite{AicSex71}
obtained a specific impulsive {\it pp\,}-wave spacetime by boosting a
spherically symmetric point source (described by the Schwarzschild
metric) to the speed of light. The same approach has subsequently been
used by a number of authors who boosted other solutions of the Kerr-Newman
or the Weyl families \cite{KerNew}. Penrose~\cite{Pen72}
 has presented yet another geometrical method for the
construction of  general impulsive {\it pp\,}-waves in a Minkowski background.
This is based on cutting the spacetime along a null plane and then
re-attaching the two pieces with a suitable warp, prescribed by the
Penrose junction conditions. These spacetimes are often presented
in terms of a metric which contains the Dirac delta
explicitly. However, it is possible to find a
coordinate system for impulsive waves, in which the metric is
continuous. For the Aichelburg-Sexl solution this has been done by D'Eath
\cite{DE78} (and used for analysis of ultrarelativistic encounters of black
holes), a continuous system for general impulsive {\it
pp\,}-wave spacetimes has been presented in \cite{AiBa97,PodVes98b}.

All the above methods can also be used to construct nonexpanding
impulsive waves in backgrounds with nonvanishing cosmological constant
$\Lambda$. Somewhat surprisingly, this has explicitly been done only
recently. It has been demonstrated \cite{Podol98} that such spacetimes can
be obtained as a distributional limit of sandwich waves of the Kundt class
$KN(\Lambda)$ of solutions~\cite{ORR}. In 1993 Hotta and Tanaka,
following the Aichelburg and Sexl approach, boosted the Schwarzschild-(anti--)de~Sitter
 spacetime to the speed of light \cite{HotTan93}. They obtained
a specific solution which represents an impulsive 2-sphere generated by a
pair of two null particles at the poles, which propagates through the
de~Sitter universe. In the anti--de~Sitter universe the impulsive wavefront
is hyperboloidal and  is generated by a single particle moving with
the speed of light \cite{PodGri97}. A general class of these spacetimes has
been presented in \cite{PodGri98}. It has also been demonstrated that
impulsive pure gravitational waves of this type are generated by null
particles with  arbitrary multipole structures. Nonexpanding
impulsive waves in the (anti--)de~Sitter spacetime can alternatively be
obtained by the Penrose ``cut and paste'' method \cite{PodGri99},
or equivalently using the ``shift function'' method \cite{Sfet}.
In \cite{PodGri99} both distributional and continuous metric forms of these
solutions have been presented, in which the limit $\Lambda\to0$
(resulting in impulsive {\it pp\,}-waves) can be explicitly performed.

Various properties of impulsive {\it pp\,}-waves in Minkowski space
have been studied. In particular, symmetries (which form a ``richer''
structure than those for waves with a smooth profile) have been
investigated in \cite{AichBal2,AiBa97}.
Geodesics have been discussed in several works \cite{DT}.
However, since the corresponding equations of geodesic and geodesic
deviation contain highly singular products of distributions, the
advanced framework of Colombeau algebras of generalized functions
had to be employed to solve these equations in a mathematically
rigorous sense \cite{Stein}.

It is the purpose of this paper to present principal results
concerning symmetries (section~3) and behaviour of geodesics
(sections~4 and 5) in spacetimes which describe  nonexpanding
impulsive waves in the (anti--)de~Sitter background.

\section{General forms of the solutions}

It has previously been shown \cite{PodGri98,PodGri99} that a
complete class of nonexpanding impulsive waves in spacetimes with
a nonvanishing cosmological constant $\Lambda$ can conveniently be
written using a 5-dimensional formalism as metrics
\begin{equation}
\d s^2=H(Z_2,Z_3,Z_4)\,\delta( U)\,\d U^2
   -2\,\d U\d V + \d{Z_2}^2 +\d{Z_3}^2+\epsilon\,\d{Z_4}^2\ ,
\label{general}
\end{equation}
on the 4-dimensional hyperboloid
\begin{equation}
-2\, U V + {Z_2}^2+{Z_3}^2+\epsilon{Z_4}^2=\epsilon a^2\ , \qquad
a=\sqrt{3/(\epsilon\Lambda)}\ ,
\label{hyperb}
\end{equation}
where \ $ U={1\over\sqrt2}(Z_0+Z_1)\,$\  and \ $ V={1\over\sqrt2}(Z_0-Z_1)\,$.
The wave is absent when  $H=0$, in which case (\ref{general}) with the constraint
(\ref{hyperb}) reduces to the well-known form of  the de~Sitter space (for  $\Lambda>0$
and $\epsilon=1$) or the anti--de~Sitter space (for $\Lambda<0$ and $\epsilon=-1$).
For a nontrivial $H$, the solution represents an impulsive wave propagating in the
(anti--)de~Sitter universe. The impulse is located on  the null hypersurface
$ U=0$ given by
\be
 {Z_2}^2+{Z_3}^2+\epsilon{Z_4}^2=\epsilon a^2\ .\label{surface}
\ee
This is a nonexpanding 2-sphere in
the de~Sitter universe, or a hyperboloidal 2-surface in the anti--de~Sitter universe
\cite{PodGri97} which can be parametrized as
\be
Z_2=a\sqrt{\epsilon(1-z^2)}\,\cos\phi\ ,\quad
Z_3=a\sqrt{\epsilon(1-z^2)}\,\sin\phi\ ,\quad
Z_4=a\, z\ .  \label{param}
\ee

In general, the above metrics describe impulsive gravitational waves
and/or impulses of null matter. Pure gravitational waves occur when
the {\it vacuum} field equation \be
(\Delta+{\textstyle{2\over3}}\Lambda)H=0\ , \label{vacuum}
\ee
where $\Delta={1\over3}\Lambda\{\partial_z[(1-z^2)\partial_z]+(1-z^2)^{-1}
\partial_\phi\partial_\phi)\}$ is the Laplacian on the impulsive surface,
is satisfied \cite{HorItz99,PodGri99,Sfet}. Non-trivial
solutions of (\ref{vacuum}) can be written as
 \begin{equation}
 H(z,\phi)= \sum_{m=0}^\infty b_mH_m(z,\phi)
= \sum_{m=0}^\infty b_mQ^m_1(z)\cos[m(\phi-\phi_m)]\ , \label{E4.5}
\end{equation}
 where $b_m$ and $\phi_m$ are constants representing an arbitrary
amplitude and phase of each component, and $Q^m_1(z)$ are associated
Legendre functions of the second kind generated by the relation
$Q^m_1(z)=(-\epsilon)^m|1-z^2|^{m/2}(\d^mQ_1(z)/\d z^m)$.
The first term (for $m=0$) alone
\be
Q_1(z)=Q^0_1(z)={z\over2}\log\left|1+z\over 1-z\right|-1\ ,  \label{HT}
\ee
gives the simplest (axially symmetric)
solution found by Hotta and Tanaka \cite{HotTan93}. The components $H_m$
describe impulsive gravitational waves generated by null point sources with an
$m$-pole structure localized at the singularities $z=\pm1$ on
the wave-front, see  \cite{PodGri98}.

Various 4-dimensional parametrizations of the solutions
(\ref{general}), (\ref{hyperb}) are known. For example,
\begin{eqnarray}
 &&Z_2 = x/\Omega\ , \qquad Z_3 = y/\Omega\ , \nonumber\\
 &&Z_4 = a\, (2/\Omega-1)\ , \label{Zcoords}\\
 &&  U = u/\Omega \ , \qquad\   V = v/\Omega \ , \nonumber
 \end{eqnarray}
with
\begin{equation}
\zeta={\textstyle{1\over\sqrt2}}(x+{\rm i}y)\ ,\qquad
\Omega= 1+{\textstyle{1\over6}}\Lambda(\zeta\bar\zeta-uv)\ , \label{zeta}
\end{equation}
brings the metric to the form
\begin{equation}
\d s^2= {2\tilde H(\zeta,\bar\zeta)\,\delta(u)\,\d u^2
    -2\,\d u\,\d v + 2\,\d\zeta\,\d\bar\zeta
\over [1+{1\over6}\Lambda(\zeta\bar\zeta-uv)]^2}\ ,
 \label{confpp}
 \end{equation}
in which $2\tilde H(\zeta,\bar\zeta) = {\textstyle(1+{1\over6}\,\Lambda\,\zeta\,
\bar\zeta)}\, H(\zeta,\bar\zeta)$. The solutions (\ref{confpp}) are written in unified
form both for $\Lambda>0$ and $\Lambda<0$. For $\Lambda=0$ these  reduce
to the well-known distributional (Brinkmann) form of the {\it pp\,}-waves
in Minkowski space.  The impulsive wave (in any spacetime
of constant curvature) is  located on the wavefront $u=0$.

Another form of the solutions is obtained
by performing the transformation
 \begin{eqnarray}
u&=&\U\ , \nonumber\\
v&=&\V+\tilde H\,\Theta(\U)+\tilde H_{Z} \tilde H_{\bar Z}\,\Theta(\U)\,\U\ ,
    \label{trans}\\
\zeta&=&Z+\tilde H_{\bar Z}\,\Theta(\U)\,\U\ , \nonumber
 \end{eqnarray}
where $\Theta(\U)$ is the Heaviside step function. This transformation
is discontinuous on $u=0$ in exact correspondence with the junction conditions
prescribed by Penrose in the ``cut and paste'' method \cite{Pen72}.
The resulting line element,
 \begin{equation}
\d s^2= { -2\,\d\, \U\,\d \V + 2\,|\d Z
  +\Theta(\U)\,\U\,(\tilde H_{Z\bar Z}\,\d Z+\tilde H_{\bar Z\bar Z}\,\d\bar Z)|^2
\over [1+{1\over6}\Lambda(Z\bar Z-\U \V+\tilde G\, \Theta(\U)\,\U)]^2}\ ,
 \label{cont}
 \end{equation}
where $\tilde G=Z\tilde H_Z+\bar Z\tilde H_{\bar Z}-\tilde H$,
is  continuous  across the null hypersurface $\U=0=u$.
However,  discontinuities in  derivatives of the metric in
general yield components in the curvature and the Weyl tensors
proportional to the Dirac $\delta$.  Note finally that for $\Lambda=0$
the metric (\ref{cont}) reduces to the Rosen form of impulsive
{\it pp\,}-waves  \cite{PodVes98b}.

\newpage

\section{Symmetries of the impulsive solutions}

Symmetries of impulsive {\it pp}-waves propagating in Minkowski space
were investigated by Aichelburg and Balasin \cite{AiBa97,AichBal2}.
Using the distributional form (given by (\ref{confpp}) for $\Lambda=0$)
they demonstrated that spacetimes with impulsive waves admit more symmetries
than the same class of waves with a general profile, e.g. sandwich
waves. In fact, {\it all} impulsive {\it pp\,}-waves (with {\it no
restriction} on the form of the structural function $\tilde H$)
admit at least a 3-parameter group of motions generated by the
Killing vector fields
\begin{equation}
  x\,(\partial/\partial {v})+u\,(\partial/\partial x)\ ,\quad
  y\,(\partial/\partial {v})+u\,(\partial/\partial y)\ ,\quad
  (\partial/\partial v)\ .
  \label{ppsym}
\end{equation}
Additional symmetries occur for specific forms of the structural function $\tilde H$,
see  \cite{AichBal2}. It is the purpose of this part of our contribution to
investigate symmetries of exact nonexpanding impulsive waves which propagate in
the de~Sitter or anti--de~Sitter backgrounds.

Spacetimes of constant curvature admit a 10-parameter group of motions. It is
convenient to study symmetries of the (anti--)de~Sitter spacetime in a
5-dimensional formalism using the coordinates introduced in (\ref{general}),
(\ref{hyperb}). Obviously, this metric representation (with $H=0$)
is invariant under the $SO(1,4)$ or  $SO(2,3)$ group of
transformations for $\Lambda>0$ or $\Lambda<0$, respectively.
In the first case, these are 3 spatial rotations, 1 boost and 6 null
rotations, in the second case 1 spatial rotation, 3 boosts and 6 null rotations
generated by the following Killing vector fields:
\begin{eqnarray}
&&  Z_2\,(\partial/\partial {Z_3})-Z_3\,(\partial/\partial {Z_2})
     \label{rot}\\
&&  Z_4\,(\partial/\partial {Z_i})-\epsilon Z_i\,(\partial/\partial {Z_4})
   \label{boost1}\\
&&   U\,(\partial/\partial { U})- V\,(\partial/\partial { V})
    \label{boost2}\\
&&  Z_i\,(\partial/\partial { V})+ U\,(\partial/\partial {Z_i})
   \label{nullV}\\
&&  \epsilon Z_4\,(\partial/\partial { V})+ U\,(\partial/\partial {Z_4})
     \label{nulltV}\\
&&  Z_i\,(\partial/\partial { U})+ V\,(\partial/\partial {Z_i})
   \label{nullU}\\
&&  \epsilon Z_4\,(\partial/\partial { U})+V\,(\partial/\partial {Z_4})
    \label{nulltU}
\end{eqnarray}
in which $i=2,3$. Finite transformations corresponding to (\ref{rot})-(\ref{boost2})
have well-known forms. Null rotations generated by (\ref{nullV})  are
\begin{eqnarray}
&&Z_2'=Z_2+b\, U\ ,  \nonumber\\
&&Z_3'=Z_3\ ,\quad Z_4'=Z_4\ ,\quad  U'= U\ ,
              \label{finnullU}\\
&& V'= V+b\,Z_2+{\textstyle{1\over2}}b^2\, U\ ,  \nonumber
\end{eqnarray}
for $i=2$. For $i=3$, they are obtained  by interchanging $Z_2\leftrightarrow Z_3$.
Null rotations corresponding to (\ref{nulltV}) are
\begin{eqnarray}
&&Z_4'=Z_4+b\, U\ ,  \nonumber\\
&&Z_2'=Z_2\ ,\quad Z_3'=Z_3\ ,\quad  U'= U\ ,
              \label{finnulltU}\\
&& V'= V+\epsilon b\,Z_4+\epsilon{\textstyle{1\over2}} b^2\, U\ .  \nonumber
\end{eqnarray}
The symmetries (\ref{nullU}), (\ref{nulltU}) are obtained from
(\ref{finnullU}), (\ref{finnulltU}) by interchanging $ U\leftrightarrow V$.

Now, let us consider impulsive waves in the (anti--)de~Sitter universe
of the form (\ref{general}) with an  arbitrary $H$. Since all
these solutions reduce to the constant-curvature spacetimes everywhere except
on the null hypersurface $\,U=0$, it is natural to look for the symmetries
of the {\it complete} solutions (including the impulse localized at $ U=0$)
among the group of transformations generated by (\ref{rot})-(\ref{nulltU}).
It is straightforward to show that there are {\it at least
three Killing vector fields}, even for a {\it general} $H$,
namely the null rotations (\ref{nullV}), (\ref{nulltV}).
Thus the spacetimes representing nonexpanding impulsive waves in
the (anti--)de~Sitter universe in general admit three symmetries of the form
(\ref{finnullU}), (\ref{finnulltU}).

Using (\ref{Zcoords}) and the inverse relations,
$x=\Omega\,Z_2$, $y=\Omega\,Z_3$,
$u=\Omega\, U$, $v=\Omega\, V$,
in which the conformal factor is  $\Omega=2a/(Z_4+a)$,
we express the three Killing vectors in the
coordinates of the metric form (\ref{confpp})
\begin{eqnarray}
&&x\,(\partial/\partial v)+u\,(\partial/\partial x)\ ,  \nonumber\\
&&y\,(\partial/\partial v)+u\,(\partial/\partial y)\ ,  \label{Killuv}\\
&&[1-{\textstyle{1\over12}\Lambda}(x^2+y^2)]\,(\partial/\partial v)
-{\textstyle{1\over6}\Lambda}\,u\left[ x\,(\partial/\partial x)
+y\,(\partial/\partial y)+u\,(\partial/\partial u)\right] \ . \nonumber
\end{eqnarray}
Thus, the corresponding transformations which leave the metric
(\ref{confpp}) unchanged  are
\begin{eqnarray}
1.&&x'=x+b_1u\ ,\quad y'=y\ , \nonumber\\
&&u'=u\ , \quad v'=v+b_1x+{\textstyle{1\over2}}b_1^2u\ . \nonumber\\
2.&&x'=x\ ,\quad y'=y+b_2u\ ,\nonumber\\
&&u'=u\ , \quad v'=v+b_2y+{\textstyle{1\over2}}b_2^2u\ . \label{sym}\\
3.&&x'={x\over{1+{\textstyle{1\over6}}\Lambda b_3\,u}}\ ,\quad
y'={y\over{1+{\textstyle{1\over6}}\Lambda b_3\,u}}\ ,\quad
u'={u\over{1+{\textstyle{1\over6}}\Lambda b_3\,u}} \ , \nonumber\\
&&v'={v+{\textstyle{1\over6}}\Lambda b_3^2\,u+b_3
[ 1-{\textstyle{1\over12}}\Lambda(x^2+y^2-2uv) ]
\over{1+{\textstyle{1\over6}}\Lambda b_3\,u}}\ . \nonumber
\end{eqnarray}
Obviously, the Killing vector fields (\ref{Killuv})
are exactly the vectors (\ref{ppsym}) found previously by
Aichelburg and Balasin \cite{AiBa97,AichBal2} for the case $\Lambda=0$.
In fact, the first two transformations (\ref{sym}) are the same as those
of impulsive {\it pp}-waves, and the third reduces to
$v'=v+b_3$ in the Minkowski background.

Other symmetries arise for some specific forms of the structural
function $H$. For example, in the case of the Hotta-Tanaka
solution given by (\ref{HT}), $H$ only depends on
$Z_4=\sqrt{a^2-\epsilon(Z_2^2+Z_3^2)}$. Thus, there
is an additional (fourth) axial symmetry, namely
the rotation generated by (\ref{rot})
Again, this is analogous to the axially symmetric Aichelburg-Sexl
solution for $\Lambda=0$.

\section{Geodesics}

In this section we explicitly derive all geodesics in spacetimes with
nonexpanding impulsive waves and nonvanishing cosmological constant.
Again, it is useful to employ the 5-dimensional formalism. The spacetimes
to be investigated can be understood as 4-dimensional submanifolds
$\H$ of the manifold $\M$ which is a 5-dimensional {\it pp\,}-wave
(\ref{general}), such that $\H$ is given by the constraint (\ref{hyperb}).
It is well-known (see, e.g.~\cite{O'Neil})
that a curve with tangent $\T$ in $\M$ lying also on $\H$ is a
{\it geodesic in} $\H$ if and only if its $\M$-acceleration,
$\A=\nabla_{\T}\T$, is everywhere normal to $\H$, i.e. its
$\H$-acceleration vanishes. If $\N$ is the normal vector
to $\H$ satisfying $\N\cdot\N=\epsilon$ (where the dot is the
scalar product in $\M$), the above condition $\A\sim\N$
using $\N\cdot\T=0$ gives
\be
\nabla_{\T}\T =-\epsilon\,(\T\cdot\nabla_{\T}\N)\,\N \ .
\label{cond}
\ee
In case of (\ref{general}), (\ref{hyperb}) the above vectors have the
contravariant components
\be
\T=( \dot{ U},\dot{ V},\dot{Z_2},\dot{Z_3},\dot{Z_4})\ ,
\quad \N=a^{-1}( U, V,Z_2,Z_3,Z_4)\ .\label{TN}
\ee
Note that throughout the paper we apply the distributional identity
$U\delta(U)=0$. The nonzero Christoffel symbols for (\ref{general}) are
$\Gamma^ V_{\; U U}=-{1\over2}H\delta'( U)$,
$\Gamma^ V_{\; U p}=-{1\over2}H_{,p}\,\delta( U)$,
$\Gamma^i_{\; U U}=-\frac{1}{2}H_{,i}\delta( U)$,
$\Gamma^4_{\; U U}=-\frac{1}{2}\epsilon H_{,4}\delta( U)$,
in which we introduced the notation
\be
p=2,3,4\ \,\qquad   i=2,3\ .
\ee
Thus,  $\T\cdot\nabla_{\T}\N=a^{-1}[e+{\textstyle\frac{1}{2}}
  G\delta( U)\,\dot{ U}^2]$,
where \ $e\equiv\T\cdot\T=0,-1,+1$\  for null, timelike and spacelike geodesics,
respectively, and $G=Z_p\,H_{,p}-H$ (summation convention is used).
Consequently, the equations (\ref{cond}) for geodesics in
spacetimes with nonexpanding impulsive waves (\ref{general}),
(\ref{hyperb}) are
\beqn
\ddot{ U}&=&-{\textstyle\frac{1}{3}}\Lambda\, U\,e\ ,  \nonumber \\
\ddot{ V}-{\textstyle\frac{1}{2}}H\,\delta'( U)\,\dot{ U}^2-
 H_{,p}\,\dot{Z_p}\,\delta( U)\,\dot{ U}
 &=& -{\textstyle\frac{1}{3}}\Lambda\, V\left[e+{\textstyle\frac{1}{2}}
  G\,\delta( U)\,\dot{ U}^2\right]\ ,  \label{5-geodeqs} \\
\ddot{Z_i}-{\textstyle\frac{1}{2}}
  H_{,i}\,\delta( U)\,\dot{ U}^2&=&
 -{\textstyle\frac{1}{3}}\Lambda\,Z_i\left[e+{\textstyle\frac{1}{2}}
 G\,\delta( U)\,\dot{ U}^2\right]\ ,  \nonumber\\
\ddot{Z_4}-{\textstyle\frac{1}{2}}\epsilon
  H_{,4}\,\delta( U)\,\dot{ U}^2&=&
 -{\textstyle\frac{1}{3}}\Lambda\,Z_4\!\left[e+{\textstyle\frac{1}{2}}
 G\,\delta( U)\,\dot{ U}^2\right]\ . \nonumber
 \eeqn
It is obvious that for an arbitrary $H$ and any  $\,U\not=0$ (i.e.
everywhere in front of and behind the impulse), these geodesic equations
are simple and reduce to those in ``pure'' (anti--)de~Sitter
spacetime (for which $H=0$). In fact, all these equations have the same
(decoupled) form, $\ddot X=-\epsilon e\, a^{-2}\,X$, where  $X$ stands
for \ $ U$, $ V$, $Z_i$, or $Z_4$. Thus, for $\,U\not=0$ the geodesics are
\beqn
&& X=X^0+\dot X^0\,\tau\ ,
  \hskip33.8mm\hbox{when}\quad  \epsilon e=0\ , \nonumber\\
&& X=X^0\cosh(\tau/a)+a\dot X^0\sinh(\tau/a)\ ,
  \hskip1mm\hbox{when}\quad  \epsilon e<0\ , \label{g2}\\
&& X=X^0\cos(\tau/a)+a\dot X^0\sin(\tau/a)\ ,
  \hskip5mm\hbox{when}\quad  \epsilon e>0\ . \nonumber
\eeqn
Recall that the first relation in (\ref{g2}) describes null geodesics, the second
represents timelike geodesics in the de~Sitter space or spacelike geodesics
in the anti--de~Sitter space, whereas the third line in (\ref{g2}) corresponds to
spacelike/timelike geodesics in the de~Sitter/anti--de~Sitter space,
respectively. In the above equations,  $\tau$ is an affine parameter,
and $X^0$, $\dot X^0$ are constants of integration. These ten
constants are constrained by the following three conditions
\beqn
-2\,\dot{ U}^{\,0}\dot{ V}^{\,0}+(\dot{Z}_2^0)^2+(\dot{Z}_3^0)^2+\epsilon(\dot{Z}_4^0)^2&=&e
 \ , \label{con1}\\
-2\, U^{\,0} V^{\,0}+(Z_2^0)^2+(Z_3^0)^2+\epsilon(Z_4^0)^2&=&\epsilon a^2
 \ , \label{con2}\\
- U^{\,0}\dot{ V}^{\,0}-\dot{ U}^{\,0} V^{\,0}+Z_2^0\dot{Z}_2^0+Z_3^0\dot{Z}_3^0+\epsilon Z_4^0\dot{Z}_4^0&=&0
  \ \label{con3}.
\eeqn
The first is the normalization of the affine parameter, the equation (\ref{con2}) follows from
(\ref{hyperb}), and (\ref{con3})  from its derivative.

Let us first observe that the equation (\ref{5-geodeqs}) for $ U$ is
decoupled and does not involve any distributional term. The solution,
which is {\it everywhere} given by (\ref{g2}), is a {\it smooth} function.
Using a freedom in the choice of the affine parameter
$\tau\to \dot\tau_0\tau+\tau_0$ for $e=0$, and $\tau\to \tau+\tau_0$ for
$e\not=0$\ ($\dot\tau_0$, $\tau_0$ are constants), we can simplify
$ U$ to the form
\be
  U=\tau \ , \qquad
  U=a\,\dot{ U}^0\sinh(\tau/a) \, \qquad
  U=a\,\dot{ U}^0\sin(\tau/a)\ ,  \label{U}
\ee
for $\epsilon e=0$, $\epsilon e<0$, or $\epsilon e>0$, respectively.
(The only exceptions are trivial null geodesics $ U=$const., and some
geodesics  (\ref{g2}) which do not intersect the impulse on $ U=0$.)
The relations (\ref{U}) allow us to take $\,U$ as the geodesic parameter,
in which case the three different forms of solutions (\ref{g2})
for $Z_p$ and $V$ can be written in a unified form
\be
X=X^0\,\sqrt{1-{\textstyle{1\over3}}\Lambda\,e\,{(\dot{U}^0)}^{-2}\,U^{\,2}}
+(\dot X^0/\dot U^0)\, U\ .
\label{X(U)}
\ee

Now, it is useful to employ the symmetries found in the previous section.
Without loss of generality, we can put a general solution of the geodesic
equations {\it in front} of the impulse to a much simpler form. Considering
suitable values of the parameters $b$ of the three finite  symmetry
transformations (\ref{finnullU}), (\ref{finnulltU}), we can always
achieve that the velocities $\dot Z_2^0$, $\dot Z_3^0$, $\dot Z_4^0$
vanish at \ $ U=0_-$.
Let us note that the corresponding Killing vectors commute and this permits
the symmetry transformations to be combined in an arbitrary order.
Using the constraints (\ref{con1})-(\ref{con3}) we can finally fix three of
the remaining constants.  Since $ U^0=0$ and $\dot Z_p^0=0$, equation
(\ref{con3}) implies $ V^0=0$. The relation (\ref{con1}) gives
$\dot V^0=-{1\over2}(e/\dot U^0)$. Thus, without loss of generality,
any geodesic (\ref{X(U)}) in front of the impulse can be written in the form
\be
 Z_p( U)=
  Z_p^0\,\sqrt{1-{\textstyle{1\over3}}\Lambda\,e\,{(\dot{ U}^0)}^{-2}
   \,U^{\,2}}\ , \qquad
  V( U) =-{\textstyle{1\over2}}\,e\,{(\dot U^0)}^{-2}\,U\ ,
 \label{geod<0}
\ee
where the constants $Z_p^0$, as a consequence of (\ref{con2}), are related by
\be
\epsilon[(Z_2^0)^2+(Z_3^0)^2]=a^2-(Z_4^0)^2\ .\label{constr}
\ee

We wish to present geodesics in complete spacetimes (\ref{general})
with the impulsive wave localized on $ U=0$. Geodesics
which pass through the wave
have the {\it same} form (\ref{X(U)}), both in front of the impulse
($ U<0$) and behind it ($ U>0$). However, the constants of integration may
have {\it different values} on both sides. The only remaining problem is
to find explicit relations between these constants. This would provide
a complete solution of the geodesic equations (\ref{5-geodeqs}) in the
spacetimes studied. In other words, we have to apply appropriate junction
conditions on the impulse.

For $ U<0$ the solution is given
by (\ref{geod<0}). For $ U>0$ it must be of a general form (\ref{X(U)}).
By inspecting the character of the distributional terms in the equations
(\ref{5-geodeqs}) it is natural to assume a complete solution in the form
\beqn
 Z_p( U) & = & Z_p^0\,\sqrt{1-{\textstyle{1\over3}}\Lambda\,e\,{(\dot{ U}^0)}
   ^{-2}\, U^{\,2}}  \ +\ A_p\,\Theta( U)\, U \ , \label{gen} \\
  V( U) & = &-{\textstyle{1\over2}}\,e\,{(\dot U^0)}^{-2}\, U
  \ +\ B\,\Theta( U)\,\sqrt{1-{\textstyle{1\over3}}\Lambda\,e\,{(\dot{ U}^0)}
  ^{-2}\, U^{\,2}} \ +\ C\,\Theta( U)\, U\ , \nonumber
\eeqn
where $A_p$, $B$, $C$ are suitable constants to be determined. In the above
equation, $Z_p( U)$ is continuous on \ $ U=0$ but not $C^1$ in such
a way that $\ddot{Z_p}$ contains the Dirac delta. This is consistent
with the equations (\ref{5-geodeqs}) for $Z_p$. Inserting this ansatz
into the geodesic equations~(\ref{5-geodeqs}) and  considering the
distributional identities
$f( U)\,\delta( U)=f(0)\,\delta( U)$,
$f( U)\,\delta'( U)=f(0)\,\delta'( U)-f'(0)\,\delta( U)$,
one obtains
\beqn
 &&A_i = {\textstyle\frac{1}{2}}\left[H_{,i}(0)
  - {\textstyle\frac{1}{3}}\Lambda\,Z_i^0G(0)\right]\ , \quad
 A_4  =  {\textstyle\frac{1}{2}}\left[\epsilon H_{,4}(0)
  - {\textstyle\frac{1}{3}}\Lambda\,Z_4^0G(0)\right]\ , \nonumber \\
 &&B \,= {\textstyle\frac{1}{2}}H(0)\ ,   \label{sol_BCA} \\
 &&C \,= {\textstyle\frac{1}{8}}\left[
   H_{,2}^2(0)+H_{,3}^2(0)+\epsilon H_{,4}^2(0)+
   {\textstyle\frac{1}{3}}\Lambda\,H^2(0)
  -{\textstyle\frac{1}{3}}\Lambda\left(Z_p^0\,H_{,p}(0)\right)^2\right]\ ,\nonumber
\eeqn
where $f(0)\equiv f(Z_p(0))=f(Z_p^0)$, and
the constants $Z_p^0$ are again constrained by (\ref{constr}).
It can be observed that in general there is
a discontinuity in \ $ V$ and its derivative on the impulse. In fact, the
jump on \ $ U=0$ is given by $\Delta V=B={\frac{1}{2}}H(0)$, which is
in full agreement with the Penrose junction conditions
in the  ``cut and paste'' method for constructing nonexpanding impulsive
gravitational waves \cite{Pen72,PodGri99}.

Let us emphasize that the solution (\ref{gen}), (\ref{sol_BCA}) describes {\it any}
geodesic in a {\it privileged} coordinate system  such that the transverse
velocities $\dot Z_p^0$ vanish on \ $ U=0_-$. This has been  achieved by performing
null rotations (\ref{finnullU}), (\ref{finnulltU}) with suitable choice of the
parameters  $b_p=-\dot Z_p^0/\dot U^0$. Note that these values do not depend
on the  positions $Z_p^0$, so that the coordinates are well-adapted to
describe all geodesics with the same velocities. However, for a physical
interpretation of the effects of impulsive waves on a {\it set} of test particles,
it is necessary to present a {\it general} solution which, in a fixed coordinate
system, describes all geodesics at once. This general form can easily be obtained
from (\ref{gen}) by  application of the null rotations  with $b_p=\dot Z_p^0/\dot U^0$,
which reintroduce the velocities. Thus, a general solution of geodesic
equations can be written as
\beqn
 Z_p( U) & = & Z_p^0\,\sqrt{1-{\textstyle{1\over3}}\Lambda\,e\,{(\dot{ U}^0)}
   ^{-2}\, U^{\,2}} \ +\ (\dot Z_p^0/\dot U^0)\, U
   \ +\ A_p\,\Theta( U)\, U \ , \label{complete} \\
  V( U) & = &  V^0\,\sqrt{1-{\textstyle{1\over3}}\Lambda\,e\,{(\dot{ U}^0)}
   ^{-2}\, U^{\,2}} \ +\ (\dot V^0/\dot U^0)\, U
  \ +\ B\,\Theta( U)\,\sqrt{1-{\textstyle{1\over3}}\Lambda\,e\,{(\dot{ U}^0)}
  ^{-2}\, U^{\,2}}\nonumber\\
 &&\qquad\qquad
  \ +\ {(\dot{ U}^0)}^{-1}(\dot Z_i^0A_i+\epsilon \dot Z_4^0A_4)\,\Theta( U)\, U
  \ +\ C\,\Theta( U)\, U\ , \nonumber
\eeqn
where the constants are constrained by (\ref{con1})-(\ref{con3}).
Of course, for $\dot Z_p^0=0$, this reduces to the simpler form (\ref{gen}).
It can immediately be seen that the magnitude of the refraction of geodesics
in the transverse directions $Z_p$ and the jump in the longitudinal direction $V$
are totally independent of the velocity  $\dot Z_p^0$.
However, the refraction in the longitudinal direction does depend
on the velocity.

Let us close this section with few comments  on the solution (\ref{complete}).
First, for $\Lambda=0$, the above results also describe all geodesics in impulsive
{\it pp\,}-wave spacetimes, i.e. in a Minkowski universe with a gravitational and/or null
matter impulsive plane wave-front. Setting $\epsilon=0$ in  (\ref{general}), dropping
 the constraint (\ref{hyperb}),
and considering $H=H(Z_2, Z_3)$, we obtain the standard metric for
{\it pp\,}-waves. In such a case the general solution
(\ref{complete}), (\ref{sol_BCA}) takes the  explicit form of  geodesics in
impulsive {\it pp\,}-waves, as discussed previously in a number of works,
see e.g. \cite{DT}, \cite{Stein}. In fact it can be observed that, for
$\Lambda=0$, the geodesic equations (\ref{5-geodeqs}) exactly reduce
to the system which has recently been {\it rigorously} solved  by Kunzinger
and Steinbauer \cite{Stein} using  Colombeau algebras within the context
of which products of distributions can be handled.

Note that geodesics in spacetimes representing (anti--)de~Sitter universe
with nonexpanding impulsive waves have already been briefly discussed
in \cite{Sfet} using  coordinates introduced by Dray and 't Hooft \cite{DT}.
However, this approach leads not only to  $\delta$, but also $\delta^2$ terms
in the coefficients of the geodesic
equations. Similar problems occur with other coordinate systems, e.g.  (\ref{confpp}).

The main advantage of the geometrical approach presented above is that this leads
to a simpler system (\ref{5-geodeqs}) which includes  only ``weakly'' singular
terms, in particular there is no square of   $\delta$.  In addition, the
decoupled equation for $ U(\tau)$ can explicitly and globally be solved. It is
a smooth function of the affine parameter $\tau$, which permits the solutions
of the remaining equations for $V$ and $Z_p$ to be expressed in terms of $U$
(analogously to {\it pp\,}-waves).

It can be observed that with the form (\ref{complete}) of the
solution, the system (\ref{5-geodeqs}) contains   products of $\Theta$
and $\delta$ distributions only in the equation for $V$. Thus, continuous (but not $C^1$)
functions $Z_p$ are consistent solutions. On the other hand, the
solution (\ref{complete}) for $V$ has to be considered to be only
heuristic. However, similar problems appearing in the {\it pp\,}-wave case
have recently been rigorously resolved in \cite{Stein}. Moreover,
we can alternatively  replace the equation for $V$ in (\ref{5-geodeqs}) by  the
constant-norm condition,
\ $2\,\dot{ U}\,\dot{ V}=\dot{Z}_2^2+\dot{Z}_3^2 +\epsilon\dot{Z}_4^2
 +H\,\delta(\, U)\,\dot U^2-e$. In this case neither
$\Theta\,\delta$  nor $\delta'$ terms  appear. This approach leads to the
{\it same} form of the solution (\ref{gen}).

\section{Discussion of the geodesics}

It has been demonstrated above that geodesics in the spacetimes
studied are influenced by the impulsive wave in such a way that
$Z_p( U)$ are everywhere continuous functions, but there is a
jump $\Delta V={1\over\sqrt2}(\Delta Z_0-\Delta Z_1)=B$\
on\ $ U={1\over\sqrt2}(Z_0+Z_1)=0$.
Thus, for generic geodesics there is a discontinuity on the impulse,
both in time $Z_0$ and  longitudinal spatial coordinate $Z_1$,
such that $\Delta Z_0=-\Delta Z_1={\sqrt2\over4}H(0)$.

In general, there are changes in  all velocity components when
free test particles pass through the impulsive wave,  given by
$\Delta \dot Z_p=A_p\,\dot U^0$ and
$\Delta \dot  V=A_i\dot Z_i^0+\epsilon A_4\dot Z_4^0+C\,\dot U^0$
(i.e. $\Delta\dot Z_0=-\Delta\dot Z_1={1\over\sqrt2}\Delta \dot V$).
This effect leads to a refraction of trajectories.
Introducing angles $\alpha_p$ and $\beta_p$ by
$\cot\alpha_p\equiv(\d Z_p/\d U)(U=0_-)=\dot Z^0_p/\dot U_0$ and
$-\cot\beta_p\equiv(\d Z_p/\d U)(U=0_+)=\dot Z^0_p/\dot U_0+A_p\,$,
we obtain the relation $\,\cot\alpha_p+\cot\beta_p=-A_p\,$, which
is a generalization of the ``refraction formula'' for deflection
of geodesics, previously introduced for impulsive {\it pp\,}-waves
\cite{DT,PodVes99}.

The behaviour of geodesics can be visualized in suitable
sections of the 5-dimensional space. There are two natural and,
in a sense, complementary figures. The first shows the
transverse directions ($Z_2, Z_3, Z_4$) in which
trajectories are refracted only. Another section visualizes
geodesics in the ($ U, V$)-space corresponding
to the time coordinate $Z_0$ and the longitudinal spatial direction $Z_1$.
As discussed above, in such a diagram the geodesics suffer a
refraction and also a jump. Let us now investigate the behaviour of
specific  timelike and null geodesics in some detail.

\subsection{Timelike geodesics}

In this part of our contribution we investigate the effect of
nonexpanding impulsive waves on privileged families of timelike
observers in the (anti--)de~Sitter spacetimes.

We start in front of the impulse ($ U<0$) with natural comoving geodesic
observers connected to the standard global parametrization of the
{\it de~Sitter universe},
\beqn
Z_0&=&a\,\sinh(t/a)\ ,\nonumber\\
Z_1&=&a\,\cosh(t/a)\cos\chi_0\ ,\nonumber\\
Z_2&=&a\,\cosh(t/a)\sin\chi_0\sin\vartheta_0\cos\varphi_0\ ,\label{comov}\\
Z_3&=&a\,\cosh(t/a)\sin\chi_0\sin\vartheta_0\sin\varphi_0\ ,\nonumber\\
Z_4&=&a\,\cosh(t/a)\sin\chi_0\cos\vartheta_0\ ,\nonumber
\eeqn
where $t$ is the global ``cosmic time'' and $\chi_0,
\vartheta_0,\varphi_0$ are arbitrary constant parameters on the
3-sphere $S^3$, which at any slice $t=$const. has radius $\,a\cosh(t/a)$.
Considering
\be
t=\tau+t_0\ , \quad\hbox{where}\quad \sinh(t_0/a)=-\cot\chi_0\ ,
\label{shift}
\ee
in the relations (\ref{comov}), we obtain an explicit form of the
privileged family of timelike observers. For all of them
\ $ U=a\,\dot{ U}^0\sinh(\tau/a)$ with \ $\dot{ U}^0={1\over\sqrt2}
\sin\chi_0$, so that the normalization (\ref{U}) is valid.
This means  that {\it every}  observer reaches
the impulse  \ $ U=0$ at $\tau=0$, where $\tau$ is the {\it individual
 proper time}. In view of (\ref{shift}) we conclude that
different observers cross the impulse at different values of the
cosmic time $t=t_0$, which depends on $\chi_0$. Note that
$t_0=0$ for observers in the equatorial plane
$\chi_0={\pi\over2}$, whereas $t_0=-\infty$ and $t_0=+\infty$ for observers
localized at the north ($\chi_0=0$) and the south pole ($\chi_0=\pi$) of $S^3$.
Now, it follows immediately that the relations (\ref{comov}) are of the form
(\ref{X(U)}), in which the constants are
\beqn
&&
Z^0_2=a\,\sin\vartheta_0\cos\varphi_0\ ,\quad
Z^0_3=a\,\sin\vartheta_0\sin\varphi_0\ ,\quad
Z^0_4=a\,\cos\vartheta_0\ , \nonumber\\
&& \dot Z_p^0=-(\cos\chi_0/a)\, Z_p^0\ , \quad
 V^0=-\sqrt2a\,\cot\chi_0\ , \quad
\dot V^0={1+\cos^2\chi_0\over\sqrt2\,\sin\chi_0}\ .\label{con}
\eeqn
Consequently, the complete behavior of this family of timelike geodesics
for all $\, U\,$ is described by the explicit solution (\ref{complete}) with
the above values of the constant parameters. This can
be used for a physical interpretation of the influence of the impulse on
motion of these observers.
In particular, for the {\it Hotta-Tanaka solution} (\ref{HT}),
$H=b_0Q_1(z)$, we obtain using (\ref{con})
\beqn
 &&A_2 = -{b_0\over 2a}\,{\cos\varphi_0\over\sin\vartheta_0}\ ,\quad
   A_3 = -{b_0\over 2a}\,{\sin\varphi_0\over\sin\vartheta_0}\ ,\quad
   A_4 = {b_0\over 2a}\,\log\left(\cot{\vartheta_0\over 2}\right)\ , \nonumber \\
 &&B = \frac{b_0}{2}\,\left[\cos\vartheta_0\log\left(\cot{\vartheta_0\over 2}\right)-1\right]\ , \quad
   \dot Z_i^0A_i+\epsilon \dot Z_4^0A_4=-(\cos\chi_0/a)\,B   \ ,
   \label{HTcoef}\\
 &&C = {b_0^2\over 8a^2}\left[\log^2\left(\cot{\vartheta_0\over2}\right)+{1\over\sin^2\vartheta_0}\right]
\ .\nonumber
 \eeqn
Notice that the coefficients $A_p, B, C$ are independent of $\chi_0$,
and diverge for $\vartheta_0=0$, $\vartheta_0=\pi$ where the two
singular null particles generating the impulsive wave are localized.
The explicit geodesics are given by (\ref{complete}), (\ref{con}), (\ref{HTcoef}).
These demonstrate the axial
symmetry of the spacetime in the transverse directions,
$Z_2=Z\,\cos\varphi_0$, $Z_3=Z\,\sin\varphi_0$, where
\be
 Z( U) = a\,\sin\vartheta_0\,\left[
  \sqrt{1+{2\,  U^2\over a^2\sin^2\chi_0}}  \ -\
  {\sqrt2\over a}\cot\chi_0\, U\right]
   \ -\ {b_0\over 2a\sin\vartheta_0}\,\Theta( U)\, U \ . \label{Z(U)} \\
\ee
The character of  geodesics in the ``equatorial'' plane $\chi_0={\pi\over2}$
is visualized in Fig.~1. Obviously, there is a focussing effect.
In general, a ring of test particles having fixed
values of $\chi_0, \vartheta_0$ with  $\varphi_0\in[0,2\pi)$, is {\it
focused} ($Z_2=0=Z_3$) at the value of $\, U_f\,$ for which
$\,Z(\, U_f)=0\,$, i.e.
\be
 U_f=a\,/\,\sqrt{{b_0^2\over 4a^2\sin^4\vartheta_0}+
{\sqrt2\, b_0\cot\chi_0\over a\sin^2\vartheta_0}-2}\ .
\label{Uf}
\ee
The argument of the square root is quadratic in $b_0$. Thus, for each
nonsingular $\chi_0$, $\vartheta_0$, there are two values of $b_0$
such that $\, U_f=\infty$. For $|b_0|$ larger than the
corresponding roots, the value of \ $ U_f$ is finite. For smaller
values, there is a defocussing effect.

Motion in the direction $Z_4$ in  de~Sitter universe
with the Hotta-Tanaka impulse is given by
\be
 Z_4( U) = a\,\cos\vartheta_0\,\left[
  \sqrt{1+{2\,  U^2\over a^2\sin^2\chi_0}}  \ -\
  {\sqrt2\over a}\cot\chi_0\, U\right]
   \ +\ {b_0\over 2a}\log\left(\cot{\vartheta_0\over2}
   \right)\,\Theta( U)\, U \ .
\label{Z4(U)}
\ee
It can be observed that all test particles with $\vartheta_0={\pi\over2}$
stay in the physically privileged plane  $Z_4=0$ (``perpendicular'' to the
sources), not only in front on the impulse, but also behind it. In general,
in front of the impulse ($ U<0$) the  test particles (\ref{comov})
for fixed $\chi_0$ form a two-sphere $Z_2^2+Z_3^2+Z_4^2=R^2$
with radius $R=a\,\left[\sqrt{1+(2/a^2)\sin^{-2}\chi_0\, U^2}-
  (\sqrt2/a)\cot\chi_0\, U\right]=a\,\cosh(t/a)\sin\chi_0$.
As the de~Sitter universe contracts and then re-expands,
the sphere of comoving particles undergoes the same behaviour.
At $\,U=0\,$ the particles are hit by the impulse and the sphere
starts to deform. The  deformation due to the
Hotta-Tanaka impulse (in the ``equatorial'' plane $\chi_0={\pi\over2}$)
is shown in Fig.~2. The focussing of particles on $Z=0$, which
occurs at $\,U_f$ given by (\ref{Uf}), is related to the formation of
specific caustic shapes.

The geodesic motion in the time direction $Z_0$ and longitudinal
spatial direction $Z_1$ is
given by $V(U)$. For the Hotta-Tanaka spacetime this is demonstrated
in Fig.~3. Indeed,  in the above natural coordinates, the geodesics
are refracted and also broken by the impulse.

Similarly, it is possible to study geodesics in the de~Sitter universe with
an arbitrary nonexpanding impulsive wave. The impulse is
purely gravitational when the vacuum field equation (\ref{vacuum})
is satisfied. Note that this field equation and the corresponding
Weyl tensor can also be calculated using the 5-dimensional
geometrical approach, as shown in the appendix. In particular, it can be observed
from (\ref{Suu}), (\ref{Weyl}) that there exist special conformally flat
solutions with pure radiation  given by $H=H_0=\,$const. A simple explicit
form of the timelike geodesics (\ref{comov}) influenced by such an impulse is
\be
Z_2=Z\,\sin\vartheta_0\cos\varphi_0  \ ,\quad
Z_3=Z\,\sin\vartheta_0\sin\varphi_0  \ ,\quad
Z_4=Z\,\cos\vartheta_0\\ ,
\ee
where
\be
Z(U)=a\,\sqrt{1+{2\,  U^2\over a^2\sin^2\chi_0}}  \ -\
 \sqrt2\,\cot\chi_0\, U \ +\ {H_0\over 2a}\,\Theta( U)\, U \ . \label{ZH(U)}
\ee
It is obvious that there is  {\it perfect focussing} of all the test particles
with arbitrary $\vartheta_0$ and $\varphi_0$ (which initially spanned a
two-sphere) when $Z(U_f)=0$. This occurs at
$U=U_f=2a^2/\sqrt{H_0^2-4\sqrt2a\, H_0\cot\chi_0-8a^2}$. Thus, all observers
(\ref{comov}) with the same value of $\chi_0$ are hit by the above
impulse at the same time $t_0$ given by (\ref{shift}), and then
exactly focus in a {\it single event} $Z_p=0$, $V=-a^2/(2U_f)$.
The same effect has already been described by Ferrari, Pendenza and Veneziano
\cite{DT} for null geodesics influenced by impulsive {\it pp\,}-waves with constant
energy density of null matter on the wave-front.

An analogous investigation can be performed for geodesic observers in the
{\it anti--de~Sitter} universe. The privileged family of timelike geodesics
is given by
\begin{eqnarray}
 Z_0&=&-a\,\cos(t/a)\ ,  \nonumber\\
 Z_1&=&a\,\sin(t/a)\,\sinh\psi_0\cos\vartheta_0\ , \nonumber\\
 Z_2&=&a\,\sin(t/a)\,\sinh\psi_0\sin\vartheta_0\cos\varphi_0  \ ,\label{g}\\
 Z_3&=&a\,\sin(t/a)\,\sinh\psi_0\sin\vartheta_0\sin\varphi_0  \ , \nonumber\\
 Z_4&=&a\,\sin(t/a)\,\cosh\psi_0\ ,  \nonumber
\end{eqnarray}
where $\psi_0,\vartheta_0,\varphi_0$ are arbitrary constant parameters
which specify particular geodesics. These  observers move around the
anti--de~Sitter hyperboloid in closed timelike loops along  the
intersections with the planes $Z_1\sim Z_p$. Again, for
\be
t=\tau+t_0\ , \quad\hbox{where}\quad \cot(t_0/a)=\sinh\psi_0\cos\vartheta_0\ ,
\label{shifta}
\ee
the relations (\ref{g}) represents a family of timelike observers with
$\,U=a\,\dot{ U}^0\sin(\tau/a)\,$ and
$\,\dot{ U}^0=[\sqrt2\sin(t_0/a)]^{-1}\,$. Again, for all observers,
$\,U=0\,$ at $\,\tau=0\,$. The relations (\ref{g}) corresponds to
(\ref{X(U)}) when we set the constants
\beqn
&& Z^0_2=a\,\sin(t_0/a)\sinh\psi_0\sin\vartheta_0\cos\varphi_0\ ,\nonumber\\
&& Z^0_3=a\,\sin(t_0/a)\sinh\psi_0\sin\vartheta_0\sin\varphi_0\ ,\nonumber\\
&& Z^0_4=a\,\sin(t_0/a)\cosh\psi_0\ , \label{cona}\\
&& \dot Z_p^0=a^{-1}\sinh\psi_0\cos\vartheta_0\, Z_p^0\ , \nonumber\\
&& V^0=-(a/\dot U^0)\,\sinh\psi_0\cos\vartheta_0\ ,\nonumber\\
&&\dot V^0=(2\,\dot U^0)^{-1} (1-\sinh^2\psi_0\cos^2\vartheta_0)\ .\nonumber
\eeqn
The general solution (\ref{complete}) with the above choice of parameters
describes the complete behavior of this family of timelike geodesics
in the anti--de~Sitter universe with a nonexpanding wave.

In particular, geodesics in the ``equatorial plane''
$\vartheta_0={\pi\over2}$ are given by
\beqn
&&
Z^0_2=a\,\sinh\psi_0\cos\varphi_0\ ,\quad
Z^0_3=a\,\sinh\psi_0\sin\varphi_0\ ,\quad
Z^0_4=a\,\cosh\psi_0\ , \nonumber\\
&& \dot Z_p^0=0\ , \quad
 V^0=0\ , \quad
\dot U^0=\dot V^0={\textstyle{1\over\sqrt2}}\ ,\label{acon}
\eeqn
which for the  {\it Hotta-Tanaka} solution $H=b_0Q_1(z)$
in the anti--de~Sitter universe yields
\beqn
 && A_2 = -{b_0\over 2a}\,{\cos\varphi_0\over\sinh\psi_0}\ ,\quad
    A_3 = -{b_0\over 2a}\,{\sin\varphi_0\over\sinh\psi_0}\ ,\quad
    A_4 = -{b_0\over 2a}\,\log\left(\coth{\psi_0\over 2}\right)\ , \nonumber \\
 && B = \frac{b_0}{2}\,\left[\cosh\psi_0\log\left(\coth{\psi_0\over 2}\right)
   -1\right]\ ,  \label{aHTcoef}\\
 && C = -{b_0^2\over 8a^2}\left[\log^2\left(\coth{\psi_0\over2}\right)
  -{1\over\sinh^2\psi_0}\right]
\ .\nonumber
\eeqn
Discussion of the behaviour of these geodesics and corresponding figures
would be analogous to those for the Hotta-Tanaka solution with
$\Lambda>0$. However, there are some important differences. For
example, the impulsive wave-surface is not spherical but
hyperboloidal. Also, there are not two singular point sources on
the impulse but only one localized at $\psi_0=0$.

Another interesting example of nonexpanding impulsive waves in
the anti--de~Sitter universe is a nonsingular  {\it Defrise-type impulse}
given by the metric (\ref{general}) with
$H=c a^{-5}(Z_3+Z_4)^3$, $c$ is a constant. With the parametrization
\be
u={ a\,U\over Z_3+Z_4}\ , \quad
v={ a\,V\over Z_3+Z_4}\ , \quad
x={a^2\over Z_3+Z_4}\ , \quad
y={aZ_2\over Z_3+Z_4}\ , \quad
\ee
this takes the 4-dimensional form
\begin{equation}
\d s^2={a^2\over x^2}\left(
  \d x^2 +\d y^2  -2\,\d u\,\d v
 +c{\delta(u)\over x^2}\,\d u^2 \right)\ .
\label{defr}
\end{equation}
This solution represents a gravitational plus null-matter
impulse. Specific particular geodesics (\ref{complete}),
(\ref{cona}) in this spacetime for $\vartheta_0=0$ are
\beqn
&&
Z^0_i=0\ ,\quad
Z^0_4=a\ , \quad
\dot Z_i^0=0\ , \quad
\dot Z_4^0=-\sinh\psi_0\ , \nonumber\\
&& V^0=-{\sqrt2}a\,\tanh\psi_0\ , \quad
\dot U^0={\textstyle{1\over\sqrt2}}\cosh\psi_0\ ,\quad
\dot V^0={\textstyle{1\over\sqrt2}}(1-\sinh^2\psi_0)/
\cosh\psi_0 \ ,\label{adcon}
\eeqn
for which
\be
 A_2 = 0\ ,\
 A_3 = {\textstyle{3\over2}}\,ca^{-3}\ ,\
 A_4 = -{\textstyle{1\over2}}\,ca^{-3}\ , \
 B = {\textstyle{1\over2}}\,ca^{-2}  \ , \
 C = c^2a^{-6} \ .
\ee
These geodesics are continuous in the transverse directions
$x$, $y$  such that $x=a$, $y=0$ on the impulse $u=0$. There
is a discontinuity in the longitudinal direction $\Delta
v=c/(2a^2)$. This agrees with the results presented in
\cite{Pod01}, in which various properties of the Defrise sandwich
and impulsive wave spacetimes  have been analyzed.

\subsection{Null geodesics}

For an arbitrary  null geodesic, the explicit solution
(\ref{complete}) simplifies to
\beqn
Z_p( U) & = & Z_p^0 +\dot Z_p^0\, U +A_p\,\Theta( U)\, U \ , \label{null} \\
  V( U) & = &  V^0\,+\,\dot V^0\, U
  +B\,\Theta( U)
  +(\dot Z_i^0\,A_i+\epsilon \dot Z_4^0\,A_4)\,\Theta( U)\, U
  +C\,\Theta( U)\, U\ . \nonumber
\eeqn
in which \ $ U$ is an affine parameter, see (\ref{U}).
In the above coordinates, the trajectories are simple straight
lines, both in front of and behind the impulse. As in the
timelike case, at \ $ U=0$ these null geodesics are refracted and
broken (in the longitudinal direction).
One could easily study the (possible) focussing of these null  geodesics.
It is obvious from (\ref{null}) that each null particle  crosses $Z_p=0$ at
\ $ U_{f\,p}=-Z_p^0/(\dot Z_p^0+A_p)$.
This, of course,  depends on the initial position $Z_p^0$ of a particle,
its velocity parameter $\dot Z_p^0$,  and the specific solution
given by $H$. Thus, the focussing is in general astigmatic.
However (as already described above for timelike geodesics), for $H=$~const.
and particular families of observers, the focussing is exact.
This effect was previously observed for special impulsive {\it pp\,}-waves
which are conformal to the above solution. This is not
surprising since the form of all {\it null} geodesics in
conformally related spacetimes is the {\it same} (up to reparametrization).
Thus, the geodesics (\ref{null}) in the spacetime (\ref{confpp}) with
$\Lambda\not=0$  have the same form as in the spacetime  (\ref{confpp}) with $\Lambda=0$,
i.e. in impulsive {\it pp\,}-waves.

Instead of presenting a specific analysis of behaviour of null geodesics in
particular impulsive spacetimes (such as the  Hotta-Tanaka
solution (\ref{HT}), or the Defrise solution (\ref{defr})), we  finally
present an alternative and equivalent derivation of the solution (\ref{null}).
In fact, this explicit form of null geodesics can also be found using the 4-dimensional
coordinates for the (anti--)de~Sitter spacetime with
impulsive waves. It can easily be shown that there exist privileged null
geodesics given by $\V=0$ with constant values $Z=Z^0$ of the complex transverse
spatial coordinate of the {\it continuous} form of the metric (\ref{cont}).
The remaining geodesic equation, which reduces to
$\,\ddot\U+\Gamma^{\,\U}_{\>\U\U}\,\dot\U^2=0\,$,
for the initial conditions $\,\U(0)=0\,$, $\,\dot\U(0)=1\,$
yields the solution
\be
\U(\sigma)={(1+{\textstyle{1\over6}}\Lambda Z^0\bar Z^0)\,\sigma\over
1+{\textstyle{1\over6}}\Lambda [ Z^0\bar Z^0-\tilde
G_0\Theta(\sigma)\,\sigma]}\ ,
\ee
where $\sigma$ is an affine parameter, and $\tilde G_0=\tilde G(Z_0,\bar Z_0)$.
Consequently, the conformal factor introduced in (\ref{zeta}) evaluated along the
above null geodesics,
$\Omega=1+{\textstyle{1\over6}}\Lambda [ Z^0\bar Z^0+\tilde G_0\Theta(\U)\,\U\,]>0$,
 takes the form
\be
\Omega
={(1+{\textstyle{1\over6}}\Lambda Z^0\bar Z^0)^2 \over
1+{\textstyle{1\over6}}\Lambda [ Z^0\bar Z^0-\tilde
G_0\Theta(\sigma)\,\sigma]}
=(1+{\textstyle{1\over6}}\Lambda Z^0\bar Z^0){\U\over\sigma}\ .
\ee
Using the relations (\ref{Zcoords}) and (\ref{trans}) we immediately obtain
\be
\sigma=(1+{\textstyle{1\over6}}\Lambda Z^0\bar Z^0)\,U\ ,\qquad
{1\over\Omega}={1-{\textstyle{1\over6}}\Lambda \tilde G_0\Theta(U)\,U \over
  1+{\textstyle{1\over6}}\Lambda Z^0\bar Z^0}\ ,
\ee
and
\beqn
 Z_p( U) & = & Z_p^0 \ +\ A_p\,\Theta( U)\, U \ , \label{nulll} \\
  V( U) & =& B\,\Theta( U) \ +\ C\,\Theta( U)\, U\ . \nonumber
\eeqn
The constants $Z_p^0$ are
\be
Z^0_2+\hbox{i}Z^0_3={\sqrt2\,Z^0\over 1+{\textstyle{1\over6}}\Lambda Z^0\bar Z^0}
\ ,\qquad
Z^0_4=a{1-{\textstyle{1\over6}}\Lambda Z^0\bar Z^0
  \over 1+{\textstyle{1\over6}}\Lambda Z^0\bar Z^0}\ ,
\ee
and the parameters $A_p$, $B$, $C$ are given by (\ref{sol_BCA}).
The explicit solution (\ref{nulll}) is identical to the solution
obtained previously using the 5-dimensional formalism. Note again
that a general form of null geodesics (\ref{null}) is obtained
form (\ref{nulll}) by performing null rotations (\ref{finnullU}), (\ref{finnulltU})
with a suitable choice of the parameters  $b_p$.

\section{Concluding remarks}

Symmetries and geodesics in spacetimes which represent
nonexpanding impulsive waves in backgrounds with nonzero cosmological
constant were investigated. One of the main results is that the
coordinates (\ref{confpp}), (\ref{cont}), or the Dray and 't~Hooft
metric form \cite{Sfet}, are not useful for finding explicit
geodesics since the corresponding equations are complicated and
contain highly singular terms such as $\delta^2$. Instead, a geometrical
approach based on embedding of the 5-dimensional form of solutions
(\ref{general}) on the (anti--)de~Sitter hyperboloid
(\ref{hyperb}) leads to a more convenient system
(\ref{5-geodeqs}). This is similar to geodesic equations for
impulsive {\it pp\,}-waves \cite{Stein} and permits all geodesics
to be found in a simple explicit form (\ref{complete}) for arbitrary
value of the cosmological constant~$\Lambda$.

Discussion of these solutions, describing the effect of impulsive waves
on timelike and null test particles, was then presented. The
coefficients (\ref{sol_BCA}) were calculated for privileged
families of observers in de~Sitter and anti-de Sitter universe,
in particular for the axially symmetric Hotta-Tanaka impulse,
conformally flat pure radiation impulse, and the Defrise-type impulse.
These are responsible both for a refraction of trajectories and an
additional jump in the longitudinal  direction.
The focussing effect of geodesics was also described.

\section*{Acknowledgments}

The authors wish to thank Roland Steinbauer and Jerry Griffiths for
their useful comments and suggestions.
The work was supported by the grant GACR-202/99/0261 of the Czech
Republic, and the grant GAUK~141/2000  of the Charles University.
The stay of M.O. at the Institute of Theoretical Physics in
Prague in 1999 was also enabled by a scholarship of the Czech Ministry
of Education, and in 2000 by Istituto Nazionale di Fisica Nucleare.

\section*{Appendix: Field equation and the Weyl tensor}
\renewcommand{\theequation}{A-\arabic{equation}}
\setcounter{equation}{0}

The 5-dimensional formalism and embedding  enabled us to derive the
geodesic equations in a particularly convenient form. Here we demonstrate
that the same approach can also be used to derive the curvature, Ricci
and Weyl tensors. We start with the well-known Gauss equation
(see, e.g. \cite{Wald})
\be
 R_{\mu\nu\rho\sigma}=g_\mu^{\ \alpha}g_\nu^{\ \beta}g_\rho^{\ \gamma}
                g_\sigma^{\ \delta}\,{\cal R}_{\alpha\beta\gamma\delta}
             +\epsilon\,(K_{\mu\rho}K_{\nu\sigma}
              -K_{\mu\sigma}K_{\nu\rho}) \ ,  \label{Gauss}
\ee
in which ${\cal R}_{\alpha\beta\gamma\delta}$ is the curvature tensor of the
5-dimensional metric $\gamma_{\mu\nu}$ on ${\cal M}$ given by (\ref{general}),
the induced metric on ${\cal H}$ defined  by (\ref{hyperb}), with the normal vector
$\N=N^\mu\partial_\mu$, is $g_{\mu\nu}=\gamma_{\mu\nu}-\epsilon N_\mu N_\nu$, and
$K_{\mu\nu}\equiv g_\mu^{\ \kappa}\nabla_\kappa N_\nu=
a^{-1}[\,g_{\mu\nu}+{\textstyle{1\over2}}\,\delta_\mu^ U\,\delta_\nu^U\,G\,\delta( U)]$
 is the extrinsic curvature of ${\cal H}$ in  ${\cal M}$. The equation (\ref{Gauss}) allows
us to express the 4-Riemann and 4-Ricci tensors of the metric $g_{\mu\nu}$.
The only nontrivial components of the curvature tensor are
${\cal R}_{ U p\,  U q}=-{1\over2}H_{,pq}\,\delta( U)$, so that
\be
 R_{UrUs}=\textstyle{1\over 3}\Lambda(g_{UU}g_{rs}-g_{Us}g_{rU})
  +\textstyle{1\over 2}\left[\textstyle{1\over 3}\Lambda
 \, g_{rs}\,G -g_{r}^{\ p}g_s^{\ q}H_{,pq}\right]\delta(U)\ ,
 \label{Rie}
\ee
where $p,q,r,s=2,3,4$, other components have the standard
constant-curvature form,
$ R_{\mu\nu\rho\sigma}=\textstyle{1\over
3}\Lambda(g_{\mu\rho}g_{\nu\sigma}-g_{\mu\sigma}g_{\nu\rho})$. It is
obvious that, for $U \neq 0$, this is the Riemann tensor of a constant
curvature spacetime, whereas in the limit $\Lambda\rightarrow 0$ we
recover its form for impulsive $\it pp\,$-waves. Since $g^{UU}\delta(U)=0$,
it turns out that $R=4\Lambda$ and it is sufficient to
consider only the traceless part of the Ricci tensor, $S_{\mu\nu}\equiv R_{\mu\nu}-
\frac{1}{4}R\,g_{\mu\nu}$. Then, the Einstein equations read
$ S_{\mu\nu}=8\pi\,T_{\mu\nu}$. Straightforward calculation shows that the
only nonvanishing component of $S_{\mu\nu}$ is
\be
 S_{\, U U}=-{\textstyle{1\over 2}}\left[\, H_{,22}+H_{,33}+\epsilon H_{,44}-
    \textstyle{1\over 3}\Lambda\,Z_p\,Z_q\,H_{,pq}
    -\textstyle{2\over 3}\Lambda \,G \,\right]\,\delta( U) \
    .\label{Suu}
\ee
It is obvious that the spacetime is everywhere a vacuum solution, except
possibly on the impulse localized at \ $ U=0$. In general, the impulse
consists of a gravitational-wave and null matter components. Notice
that $S_{\, U U}$ is linear in $H$ so that the superposition of arbitrary
sources is also allowed. Purely gravitational impulsive waves arise when
\be
H_{,22}+H_{,33}+\epsilon H_{,44}-
    \textstyle{1\over 3}\Lambda\,Z_pZ_q\,H_{,pq}
    +\textstyle{2\over 3}\Lambda\,(H-Z_p\,H_{,p})=0\ .\label{vac5}
\ee
Considering the parametrization (\ref{param}) of the
two-dimensional impulsive wave-surface (\ref{surface}), the  equation
(\ref{vac5}) gives exactly (\ref{vacuum}), which is the form already
derived using different approaches in
\cite{HorItz99,PodGri99,Sfet}. In the limit $\Lambda= 0$, $\epsilon=0$
we recover the well-known vacuum field equation for $pp$-waves.

By means of~(\ref{Rie}), (\ref{Suu}) and the standard decomposition
of the Riemann tensor, we can calculate the Weyl tensor,
which represents  a contribution of pure gravitational waves to curvature.
The only nontrivial nonvanishing components are
\be
 C_{UrUs}=\left[-\textstyle{1\over 2}g_r^{\ p}g_s^{\ q}H_{,pq}
 +\textstyle{1\over 4}g_{rs}\left(H_{,22}+H_{,33}+\epsilon H_{,44}
 -\textstyle{1\over 3}\Lambda\,Z_p\,Z_q\,H_{,pq}\right)\right]\delta(U)\ . \label{Weyl}
\ee
 These are linear and homogeneous in the second derivatives
of $H$. It follows that $H$ which is either constant or linear in the
coordinates $Z_p$ corresponds to a {\it globally conformally flat} spacetime.
For $H=$const. one gets $T^{\, V V}=-\textstyle{1\over 24\pi}
\Lambda\,H\,\delta( U)$, the energy-momentum tensor representing
a homogeneous distribution of null matter forming the impulse.
For $H$ linear in $Z_p$, the spacetime is {\it everywhere} a conformally
flat vacuum solution, which is the (anti--)de~Sitter spacetime, i.e. a
trivial constant curvature universe with no impulse.

\newpage

\begin{figure}
\centering
\label{Figure 1}
\caption{
Refraction in the transverse directions $Z_2=Z\,\cos\varphi_0$, $Z_3=Z\,\sin\varphi_0$ of
trajectories~(\ref{Z(U)}) of timelike geodesics  by the axially symmetric Hotta-Tanaka impulse
propagating in the de~Sitter universe. The privileged observers are  comoving with fixed values
of $\chi_0={\pi\over2}$, $\vartheta_0$ and $\varphi_0$ in front of the spherical impulse.}
\end{figure}

\begin{figure}
\centering
\label{Figure 2}
\caption{
Refraction of timelike geodesics in the transverse directions $Z_4$ and $Z$ by the Hotta-Tanaka impulse
in the de~Sitter universe (dashed lines) as given by (\ref{Z4(U)}) and (\ref{Z(U)})
for $\chi_0={\pi\over2}$. Also, the deformation of a sphere
of free test particles is indicated for some values of $U$. The sphere, which is initially contracting
with the contracting de Sitter universe, starts to deform at $U=0$ by the
impulse into shapes with caustic points on the axis of symmetry $Z_4$.}
\end{figure}

\begin{figure}
\centering
\label{Figure 3}
\caption{
Behaviour of timelike geodesics $(\chi_0={\pi\over2})$ in the longitudinal direction in the Hotta-Tanaka
impulsive spacetime as given by $V(U)$ in (\ref{complete}) with (\ref{HTcoef}). In addition to refraction of trajectories, there is also a discontinuity in the
natural coordinates.}
\end{figure}

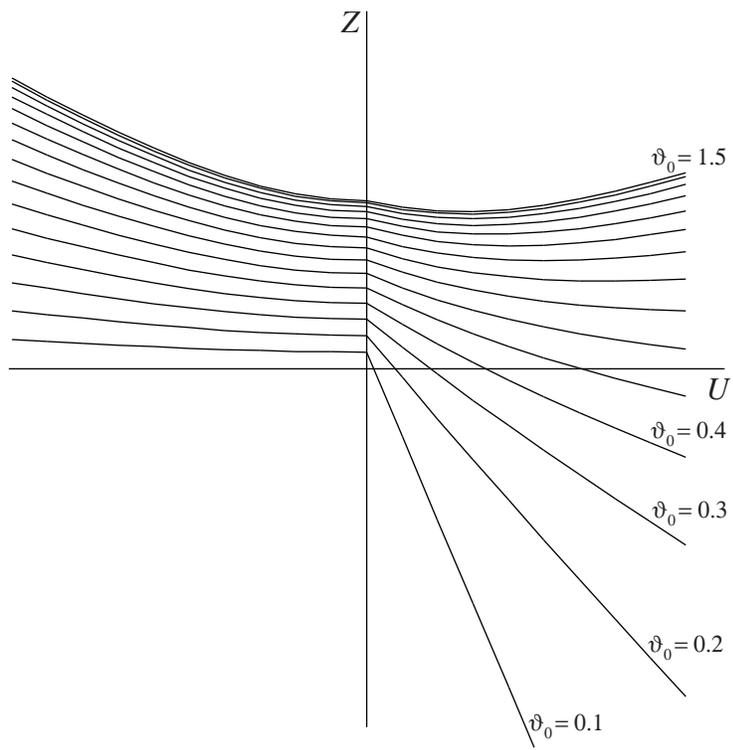
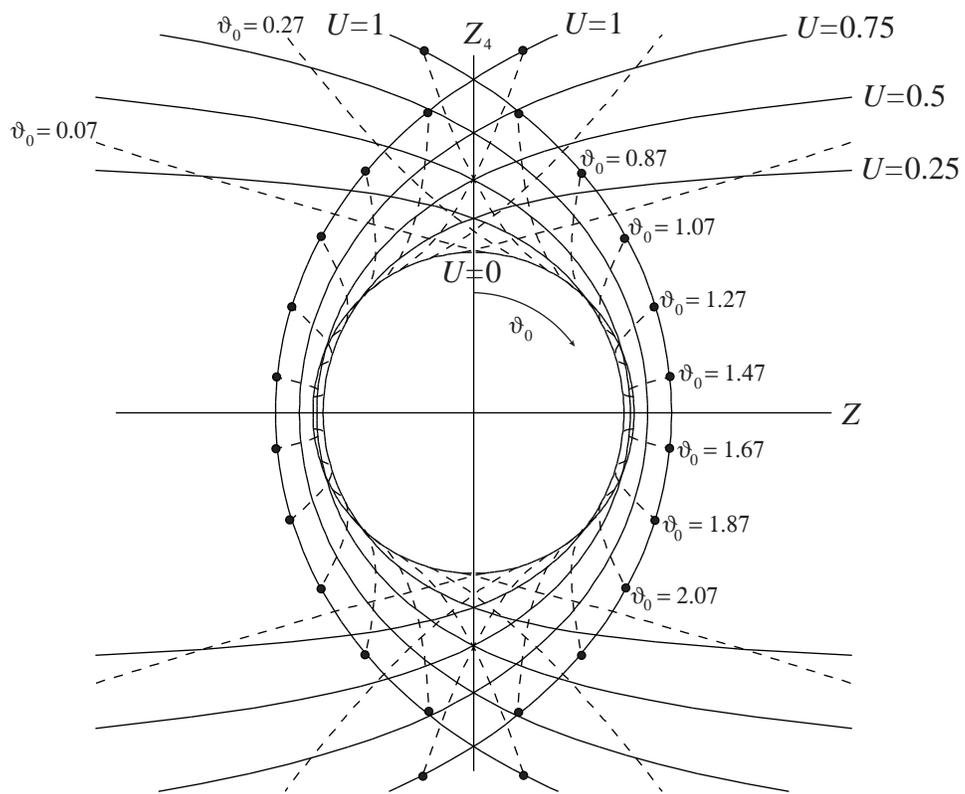
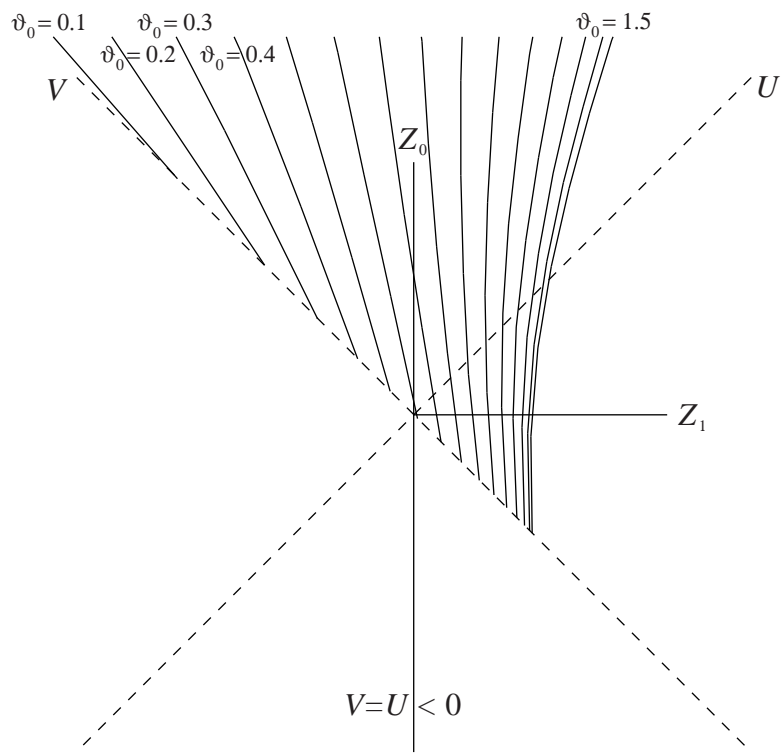
\end{document}